\documentclass[preprint2,twocolumn,times,tighten]{aastex63}

\usepackage{graphicx,times}
\usepackage{subfigure}
\newcommand{\be}{\begin{equation}}
\usepackage{threeparttable}
\usepackage{booktabs}
\newcommand{\ee}{\end{equation}}
\newcommand{\bea}{\begin{eqnarray}}
\newcommand{\eea}{\end{eqnarray}}

\usepackage{amsmath}
\usepackage{cases}
\usepackage{longtable}
\usepackage{hyperref}
\usepackage{epstopdf}
\usepackage{amsmath,bm}
\usepackage{amssymb}
\usepackage{natbib}
\usepackage{morefloats}
\usepackage{multirow}
\usepackage{array}
\usepackage{verbatim}
\usepackage{color}
\usepackage{lineno}

\begin{document}

\title{Mirror diffusion of cosmic rays in highly compressible turbulence near supernova remnants}


\author[0000-0002-0458-7828]{Siyao Xu}
\affiliation{Institute for Advanced Study, 1 Einstein Drive, Princeton, NJ 08540, USA; sxu@ias.edu
\footnote{Hubble Fellow}}

\begin{abstract}

Recent gamma-ray observations reveal inhomogeneous diffusion of cosmic rays (CRs) in the interstellar medium (ISM). 
This is expected as the diffusion of CRs depends on the properties of turbulence, which 
{can vary widely}
in the 
multi-phase ISM. 
We focus on the {mirror} diffusion 
arising in highly {compressible} turbulence in molecular clouds (MCs) around supernova remnants (SNRs), 
where the magnetic mirroring effect results in 
{significant suppression of} diffusion of CRs near {CR} sources. 
Significant 
energy loss via proton-proton interactions due to slow diffusion 
flattens the 
low-energy CR spectrum, 
while the high-energy CR spectrum is steepened due to the strong dependence of {mirror} diffusion on CR energy. 
The resulting broken power law spectrum of CRs 
{matches} well the gamma-ray spectrum observed from SNR/MC systems, e.g., 
IC443 and W44. 

\end{abstract}


\section{Introduction}

The diffusion of cosmic rays (CRs) is a fundamental physical ingredient that is 
involved in diverse astrophysical processes, 
including solar modulation 
\citep{Pot13}, 
magnetospheric shielding of extrasolar Earth-like planets
\citep{Grie15},
diffusive shock acceleration 
\citep{Axf77},
ionization in molecular clouds (MCs) and star formation 
\citep{Schlk16,Padd18},
and driving galactic winds 
\citep{Ipa75}.

Gamma-ray observations provide a powerful tool to probe the diffusion of CRs 
in the vicinity of CR sources
\citep{DiS19,Seme21}. 
Recent gamma-ray observations suggest {significant suppression of diffusion} of CRs 
with respect to the average Galactic value 
around, e.g., pulsar wind nebulae
\citep{Abey17,Evo18}
and 
supernova remnants (SNRs)
\citep{Tor10,Li12,Fun17}, 
which is also indicated by the 
enhanced ionization rate in shocked clumps and molecular clouds (MCs) around SNRs 
\citep{Ind10,Cec11}. 
In addiiton, 
steep gamma-ray spectra are found in 
SNRs interacting with MCs 
\citep{Li12,Ack13,Card14},
as well as in the Gould Belt clouds
\citep{Nero12,Bla12}.

Theoretically, with the development on fundamental theories of 
magnetohydrodynamic (MHD) turbulence in the past two decades
\citep{GS95,LV99}, 
significant progress has been made in our understanding on diffusion of CRs.
Naturally, the diffusion of CRs depends on the properties of MHD turbulence
{with which} they interact. 
Compressible MHD turbulence can be decomposed into Alfv\'{e}n, slow, and fast modes 
\citep{LG01,CL02_PRL,CL03}.
The Alfv\'{e}nic component with a Kolmogorov energy spectrum turns out to be 
inefficient in CR pitch-angle scattering due to its anisotropy
\citep{Chan00,YL02,XL20}.
It also leads to $D(E)\propto E^{-3/2}$, where $D$ is the diffusion coefficient in the direction parallel to the magnetic field, 
and $E$ is the CR energy
\citep{BYL2011,LX21},
instead of $D(E)\propto E^{1/3}$ that is expected for isotropic Kolmogorov turbulence. 
By contrast, 
fast modes, despite their relatively small energy fraction compared with Alfv\'{e}n modes
\citep{CL02_PRL,Hu21},
play a dominant role in scattering CRs especially in cold interstellar phases 
\citep{YL04,XLb18,XL20},
as well as in galaxy clusters 
\citep{Brunetti_Laz,BruLaz11},
and thus they are important for determining the diffusion of CRs. 
In the multi-phase interstellar medium (ISM),
with the varying compressibility and magnetization of the medium and driving condition of turbulence
\citep{Gae11,Laz18,Xup20,Ha21}, 
both the properties of turbulence and the turbulence-dependent diffusion of CRs 
are expected to be spatially inhomogeneous. 

More recently, 
\citet{LX21}
(henceforth LX21) 
investigated the mirroring effect in compressible MHD turbulence on CR diffusion. 
In the presence of magnetic mirrors generated by {compressible} slow and fast modes, 
$D$ corresponding to the scattering by fast modes becomes smaller 
because the range of pitch angles where 
scattering dominates over mirroring
is smaller than $[0^\circ, 90^\circ]$
\citep{CesK73,XL20}.
Moreover, 
LX21 identified a new diffusion mechanism of the CRs for which mirroring dominates over scattering. 
They found that unlike static magnetic mirrors that trap CRs 
\citep{CesK73},
due to the superdiffusion of magnetic fields induced by Alfv\'{e}nic turbulent motions
\citep{LV99,XY13,LY14}, 
CRs that interact with magnetic mirrors  
propagate diffusively along the magnetic field, 
which is termed the ``{mirror} diffusion".
As the ``mean free path" is determined by the size of {compressible} magnetic fluctuations,
the {mirror} diffusion enabled by fast modes usually has a smaller 
$D$ than that resulting from the scattering by fast modes
(LX21).

To illustrate the application of the {mirror} diffusion in the ISM, 
in this paper we focus on the diffusion of CRs in an MC-like environment 
with highly {compressible} turbulence 
(e.g., \citealt{Bal87,XuZ16,XuZ17})
and thus nonnegligible fast modes.
As the {mirror} diffusion in general 
has a smaller $D$ compared with other diffusion mechanisms in compressible MHD turbulence, 
we consider it as the dominant diffusion mechanism that governs the 
diffusion behavior of CRs near their acceleration sites. 
The MCs in the vicinity of SNRs 
serve as the target dense material for the CR protons that escape from the SNRs where they are accelerated, 
resulting in gamma-ray emission via neutral pion decay.
By formulating the distribution function of CRs that undergo both {mirror} diffusion and 
energy loss due to pion production in proton-proton (p-p) interactions,
we will be able to theoretically derive the CR energy spectrum and the corresponding gamma-ray spectrum. 
We can then test the {mirror} diffusion and its $D$ by 
comparing the theoretical expectation with the observed gamma-ray spectrum of an SNR/MC system.

This paper is organized as follows.
In Section 2, we describe the {mirror} diffusion and provide the basic formalism of its $D$. 
In Section 3, 
we present a general analysis on the distribution function and energy spectrum of CRs
with both diffusion and energy loss due to pion production taken into account. 
In Section 4, we apply the {mirror} diffusion to SNR/MC systems {and compare} the analytically derived gamma-ray spectra 
with the observed ones. 
The discussion and summary of our main results can be found 
in Section 5 and Section 6.

\section{{Mirror} diffusion of CRs}
\label{sec: bodiff}

In compressible MHD turbulence, 
the diffusion of CRs is determined by their interactions with both Alfv\'{e}nic and {compressible} components of MHD turbulence. 
The energy fraction of the {compressible} component depends on the driving mechanism of turbulence, 
the sonic Mach number $M_s = V_L /c_s$ 
and the Alfv\'{e}n Mach number $M_A = V_L/ V_A $
of the medium
\citep{CL02_PRL,Fede11,Lim20,Hu21}, where 
$V_L$ is the turbulent velocity at the injection scale $L$ of turbulence, 
$c_s$ is the sound speed,
and $V_A = B/\sqrt{4\pi\rho}$ is the Alfv\'{e}n speed with the magnetic field strength $B$ and gas density $\rho$. 
In highly {compressible} interstellar media with a large $M_s$, such as MCs
(e.g., \citealt{Bal87,Pad99,MacL04}), 
a nonnegligible
fraction of {compressible} fast and slow modes is expected 
\citep{Vaz00,Xuy21}.

Magnetic compressions induced by fast and slow modes act as magnetic mirrors. 
CRs with the gyroradius $r_g = v_\perp/\Omega$
smaller than the variation scale of the magnetic field
and the pitch-angle cosine 
\begin{equation}
    \mu < \sqrt{\frac{b_k}{B_0+b_k}}
\end{equation}
are subject to the mirror force, and they can be reflected by magnetic mirrors when they move 
along converging field lines. 
Here $v_\perp$ is the CR speed perpendicular to the magnetic field, 
$\Omega = |q| B_0/\gamma mc$ is the CR gyrofrequency
with $q$, $m$, and $\gamma$ as the electric charge, mass, and Lorentz factor of the CR particle, 
$B_0$ is the mean magnetic field strength, 
$c$ is the light speed, 
and $b_k$ is the magnetic perturbation at the wavenumber $k$
of {compressible} modes.

In the mean time, 
the shear Alfv\'{e}nic turbulent motions mix magnetic field lines and cause their superdiffusion 
in the direction perpendicular to the mean magnetic field
\citep{LV99,LY14}.
The anisotropic scaling of strong MHD turbulence is
\citep{GS95,LV99}
\begin{equation}
   l_\| = \frac{V_A}{V_\text{st}} L_\text{st}^\frac{1}{3} l_\perp^\frac{2}{3}, 
\end{equation}
where $l_\|$ and $l_\perp$ are the parallel and perpendicular sizes of a turbulent eddy measured relative to the local 
mean magnetic field
\citep{LV99,CV00},
and
$V_\text{st}$ is the turbulent velocity at the length scale $L_\text{st}$, below which the turbulent cascade is in the 
strong turbulence regime.
For sub-Alfv\'{e}nic turbulence with $M_A <1$, there are 
\citep{Lazarian06,XJL19}
\begin{equation}
  V_\text{st} = V_L M_A, ~~ L_\text{st} =l_\text{tr}= LM_A^2,  
\end{equation}
and for super-Alfv\'{e}nic turbulence with $M_A>1$, there are 
\begin{equation}
   V_\text{st} = V_A, ~~ L_\text{st} = l_A = LM_A^{-3}.
\end{equation}
It means that over a parallel distance $s$ along the turbulent magnetic field, 
the wandering field lines have spread by a distance comparable to the transverse scale corresponding to $s$, 
which is 
\begin{equation}
     \delta_\perp^2 =  L^{-1} M_A^4   s^3,   ~~ \delta_\perp < l_\text{tr}
\end{equation}
for sub-Alfv\'{e}nic turbulence, 
and 
\begin{equation}
     \delta_\perp^2 = L^{-1}M_A^3 s^3 ,   ~~\delta_\perp<l_A
\end{equation}
for super-Alfv\'{e}nic turbulence
(see \citealt{LY14} for the magnetic field diffusion in other turbulence regimes).
With the increase of $s$, the field lines are distorted by larger and larger eddies. 
As a result, $\delta_\perp^2$ has a stronger dependence on $s$ than the linear dependence, 
and the magnetic fields exhibit superdiffusion in perpendicular direction.

Due to the superdiffusion of field lines
on all length scales within the inertial range of strong MHD turbulence, 
infinitely many magnetic field lines can be stochastically advected to each point in space 
\citep{Eyink2011}. 
In the presence of both Alfv\'{e}nic and {compressible} components of MHD turbulence, 
when a CR particle is reflected by a magnetic mirror, the stochasticity of field lines makes it unable to 
trace back the same field line, and thus it
cannot be trapped and bounce back and forth between two magnetic mirrors. 
This is different from the case with the
magnetic mirrors arising from compressional MHD waves 
that was usually considered in the literature
(e.g., \citealt{CesK73}).
The resulting stochastic encounters with different magnetic mirrors lead to diffusive behavior of CRs in 
the direction parallel to the magnetic field, which is termed ``{mirror} diffusion"
(LX21).

At each $\mu$, the {mirroring effect} is dominated by the magnetic fluctuation at 
(\citealt{CesK73}; LX21)
\begin{equation}\label{eq: bostep}
   k^{-1} = L \Big(\frac{\delta B_f}{B_0}\Big)^{-4} \mu^8.
\end{equation}
Here we consider {that} fast modes dominate the {mirroring} effect, 
which is true when fast and slow modes have comparably large energy fractions 
\citep{XL20},
and adopt the scaling of fast modes
\citep{CL02_PRL}
\begin{equation}
   b_{fk} = \delta B_f (kL)^{-\frac{1}{4}}, 
\end{equation}
where $b_{fk}$ and $\delta B_f$ are the magnetic fluctuations of fast modes at $1/k$ and $L$, respectively. 
Given the $\mu$-dependent step size of {mirror} diffusion in Eq. \eqref{eq: bostep}, 
we have the corresponding $\mu$-dependent parallel diffusion coefficient (LX21)
\begin{subnumcases}
     { D_{\|,f} (\mu)=\label{eq: dufup}}
        v\mu k^{-1} = v L \Big(\frac{\delta B_f}{B_0}\Big)^{-4}  \mu^{9},  \nonumber\\
        ~~~~~~~~~~~~~~~~~~~~~~~~~~~~~\mu_\text{min,f} <\mu <\mu_c ,\\
        v\mu r_g, ~~~~~~~~~~~~~~~~~~~~~~~~~~ \mu < \mu_\text{min,f}.
\end{subnumcases}
Here 
\begin{equation}
     \mu_\text{min,f} \approx \sqrt{\frac{b_{fk} (r_g)}{B_0}}  
\end{equation}
corresponds to the minimum scale ($=r_g$) of magnetic fluctuations for the mirror bouncing
\citep{CesK73},
where $b_{fk} (r_g)$ is the magnetic fluctuation at $r_g$.
The upper limit of $\mu$ is given by the cutoff value $\mu_c$, which is determined at the balance between {mirroring} and scattering
\citep{XL20}
\begin{equation}\label{eq: mucadj}
   \mu_c \approx 1.2 \bigg[\frac{14}{\pi} \frac{\delta B_f^2}{B_0^2} \Big(\frac{v}{L\Omega}\Big)^\frac{1}{2}\bigg]^\frac{2}{11},
\end{equation}
where $v$ is the CR speed. 
The factor $1.2$ is introduced here for the analytical approximation 
to better agree with the numerical evaluation of $\mu_c$, 
and we note that $\mu_c$ cannot exceed the maximum $\mu = \sqrt{\delta B_f / (B_0+ \delta B_f)}$ for {mirroring}. 
Only at $\mu<\mu_c$, is the parallel diffusion of CRs dominated by {mirroring}. 
At $\mu>\mu_c$, the scattering by fast modes is more important in determining the parallel diffusion
\citep{XL20}.
Finally, we have the pitch-angle integrated parallel diffusion coefficient of CRs {under the mirroring effect} as 
(LX21)
\begin{equation}\label{eq: intgdp}
\begin{aligned}
   D_{\|,f} &= \int_0^{\mu_c} D_{\|,f}(\mu) d\mu \\
                  & \approx \frac{1}{10} v L \Big(\frac{\delta B_f}{B_0}\Big)^{-4} \mu_c^{10}
\end{aligned}
\end{equation}
under the assumption of an isotropic pitch angle distribution. 
Its dependence on CR energy $E$ is determined by the $E$ dependence of $\mu_c$.

\begin{figure*}[ht]
\centering
\subfigure[]{
   \includegraphics[width=8.7cm]{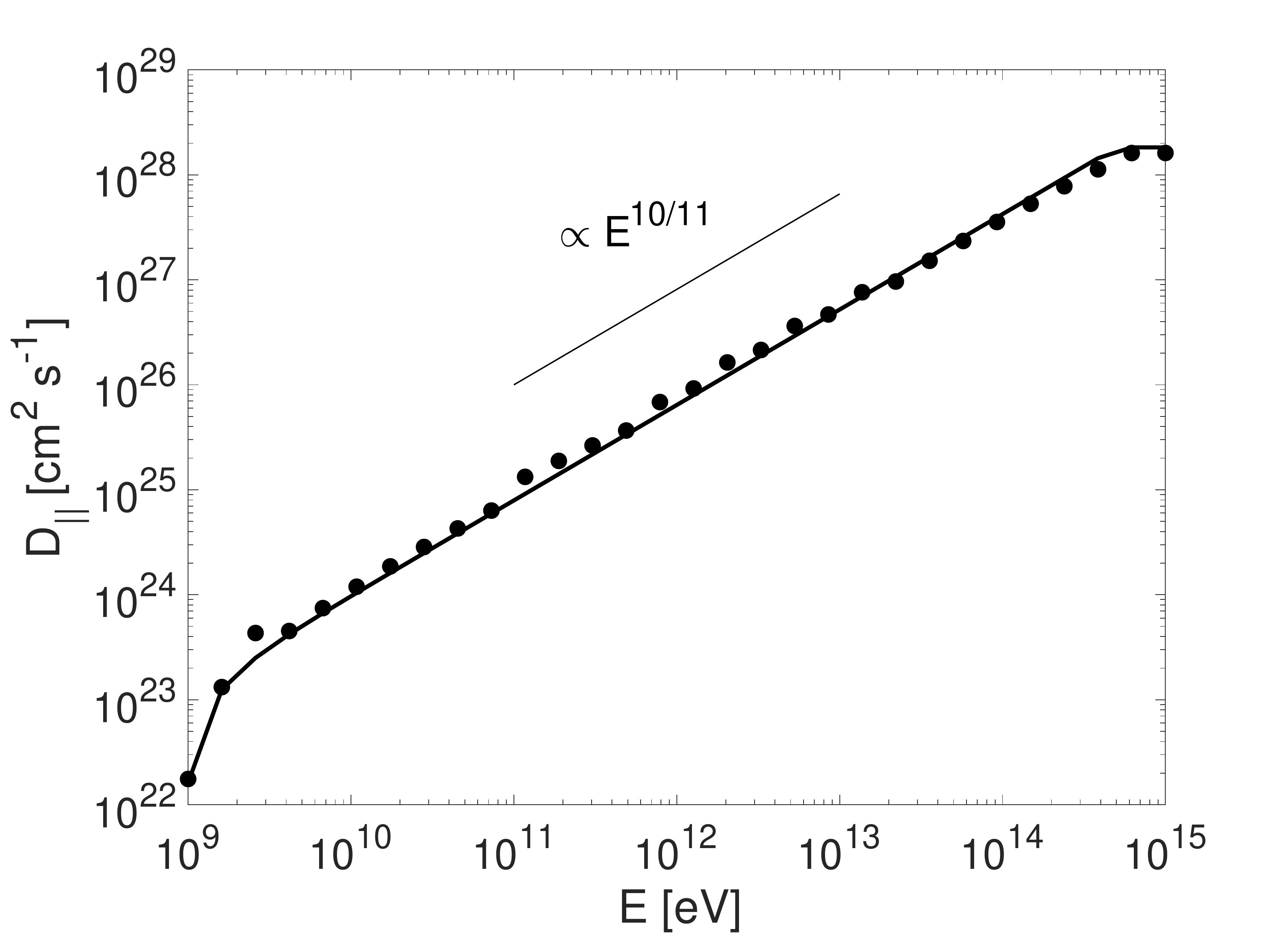}\label{fig: dbb1}}
\subfigure[]{
   \includegraphics[width=8.7cm]{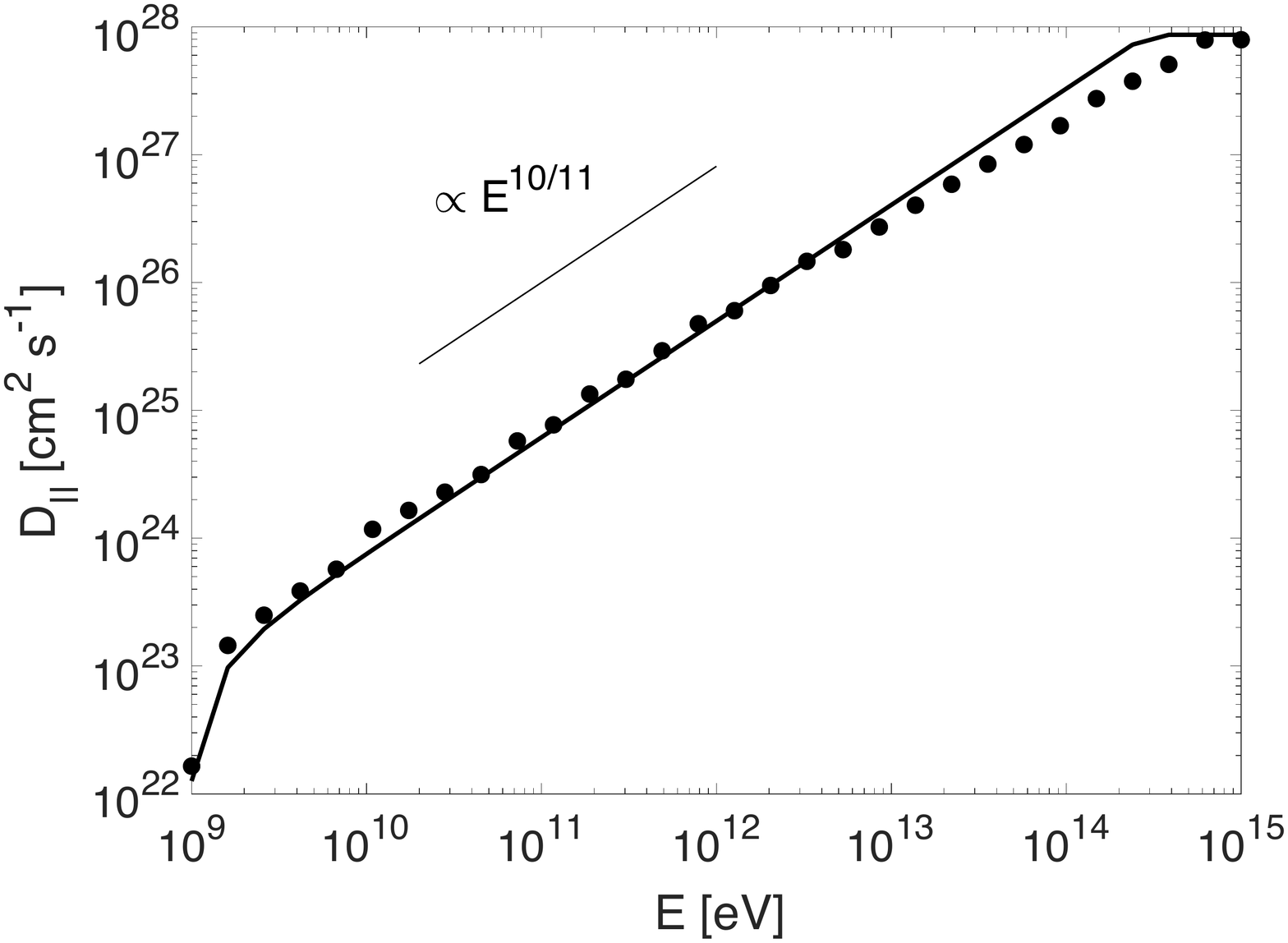}\label{fig: dbb2}}
\caption{$D_\|$ of CR protons {under the mirroring effect} in compressible MHD turbulence with 
(a) $\delta B_f = 0.5 B_0$ and 
(b) $\delta B_f = B_0$.
Dots and solid lines represent numerical calculation (see Appendix) and analytical estimate (Eqs. \eqref{eq: mucadj} and \eqref{eq: intgdp}). }
\label{fig: diffene}
\end{figure*}

In Fig. \ref{fig: diffene}, we present the numerically calculated $D_\|$
for CR protons {under the mirroring effect} in compressible MHD turbulence with $\delta B_s =0.5 B_0$ and 
$\delta B_A = B_0$ for the magnetic fluctuations of slow and Alfv\'{e}n modes at $L$
(see Appendix for other equations used). 
We consider 
$L=30~$pc and $B_0 = 70~ \mu$G for an MC-like environment.
As an example, the energy fraction of fast modes varies with $\delta B_f = 0.5 B_0$ in Fig. \ref{fig: dbb1}
and $\delta B_f = B_0$ in Fig. \ref{fig: dbb2}. 
In both cases, fast modes dominate both {mirroring} and scattering. 
The analytical estimation (Eqs. \eqref{eq: mucadj} and \eqref{eq: intgdp}) using only 
fast modes agrees well with the numerical results. 
The $E$ scaling of $D_\|$ is steep with the slope $\approx 10/11$, consistent with the analytical estimation.
We note that $D_{\|,f}$ is not very sensitive to $\delta B_f/B_0$ 
as long as it is not very small
(see LX21 for the case with a small energy fraction of fast modes).

Compared with the average Galactic value of $D$ measured near the Sun
\citep{St07,Blaa12},
\begin{equation}
    D \sim 10^{28} \Big(\frac{E}{\text{GeV}}\Big)^\frac{1}{3} \text{cm}^2\text{s}^{-1},
\end{equation}
the {mirror} diffusion in highly {compressible} turbulence in cold interstellar phases 
has a significantly smaller $D$ with a steeper $E$ scaling. 
We will show below that 
the form of $D$ is important for shaping the energy spectrum of CRs that propagate in the surrounding MCs around SNRs.


\section{Distribution function and energy spectrum of CRs}
\label{sec: discr}

The distribution function $f({\bm r}, E)$, i.e., 
the number of CR protons of energy $E$ at a position ${\bm r}$ per unit volume and unit energy interval 
in a stationary state, can be found by solving 
\citep{Syr59,Gin64}
\begin{equation}\label{eq: disfun}
     D \Delta f - \frac{\partial}{\partial E} (Sf) + Q =0. 
\end{equation}
We consider that $D$ has a power-law dependence on $E$ with the power-law index $\alpha$, 
\begin{equation}
   D=D_c E^\alpha, 
\end{equation}
and neglect its spatial dependence by assuming that 
the properties of MHD turbulence in the local environment, i.e., an MC, 
do not significantly change. 
The energy loss rate is
\begin{equation}\label{eq: enlor}
  S =   \frac{dE}{dt} = - S_c E,
\end{equation}
for which we consider the energy loss of a relativistic proton due to pion production with 
\citep{ManS94}
\begin{equation}
    S_c \approx 0.65 c n_H \sigma_\text{p,inel}, ~~ \gamma \gg 1.
\end{equation}
$n_H$ is the hydrogen density in the MC, and $\sigma_\text{p,inel}$ is the cross-section of inelastic p-p interactions. 
We further consider a point source of CRs  
\citep{Gra72}
\begin{equation}
\begin{aligned}
   &Q = Q_\epsilon E_0^{-\gamma_0},   ~~|{\bm r_0} - {\bm r_s}| < \epsilon    \\
   &Q = 0,  ~~~~~~~~~~~~~~|{\bm r_0}-{\bm r_s}| > \epsilon,
\end{aligned}
\end{equation}
where $\gamma_0$ is the power-law index of the injected CR distribution, 
${\bm r_s}$ is the source position, ${\bm r_0}$ and $E_0$ are the initial position and initial energy 
of CRs, 
and $Q_\epsilon$ and $\epsilon$ satisfy 
\begin{equation}
     \underset{\epsilon \rightarrow 0}{\text{lim}} \frac{4}{3} \pi \epsilon^3 Q_\epsilon = Q_c.
\end{equation}
For CRs injected at the SN shock, 
a spatially more extended source should be modeled, but here we adopt a 
point source for the injection of CRs to focus on the 
diffusion effect within the volume under consideration.

By introducing the variable 
\begin{equation}
   \lambda = \int_{E_0}^E \frac{D}{S} dE 
                = \frac{D_c}{S_c \alpha} (E_0^\alpha - E^\alpha), 
\end{equation}
the general solution to Eq. \eqref{eq: disfun} for infinite space is 
\citep{Syr59}
\begin{equation}\label{eq: gensol}
\begin{aligned}
&   f({\bm r},E) = \iiint d {\bm r_0} \int d E_0 Q f({\bm r},E,{\bm r_0},E_0) , \\
                  &= \frac{Q_c}{S_c E} \int_E^\infty  E_0^{-\gamma_0} \frac{1}{ (4\pi  \lambda)^\frac{3}{2} } \exp{\Bigg[-\frac{({\bm r}-{\bm r_s})^2}{4\lambda}\Bigg]} dE_0 ,
\end{aligned}
\end{equation}
where 
\begin{equation}
    f({\bm r},E,{\bm r_0},E_0) = \frac{1}{S_c E (4\pi  \lambda)^\frac{3}{2} } \exp{\Bigg[-\frac{({\bm r}-{\bm r_0})^2}{4\lambda}\Bigg]}.
\end{equation}

We see that when $\lambda \ll ({\bm r}-{\bm r_s})^2$, 
the integral over $E_0$ in Eq. \eqref{eq: gensol} is independent of $E$, and there is 
\begin{equation}\label{eq: lensc}
   f({\bm r},E)\propto E^{-1}. 
\end{equation}
This energy scaling is determined by the $E$-dependence of $S$ (Eq. \eqref{eq: enlor}). 
At the opposite limit with $\lambda \gg ({\bm r}-{\bm r_s})^2$, we introduce 
\begin{equation}\label{eq: expd}
    d^2 =   \frac{D_c}{S_c \alpha} E^\alpha .
\end{equation}
$d$ is the distance over which CRs with energy $E$ have a significant energy loss. 
So for CRs with  
\begin{equation}\label{eq: enebrk}
    E \gg E_d = \Big(\frac{d^2 S_c \alpha}{D_c}\Big)^\frac{1}{\alpha}, 
\end{equation}
their energy loss is negligible. 
By inserting Eq. \eqref{eq: expd} in Eq. \eqref{eq: gensol}, 
we approximately have 
\begin{equation}\label{eq: nocool}
  f({\bm r},E) \approx \frac{Q_c  \alpha}{ (4\pi  )^{\frac{3}{2}} D_c d (\gamma_0 + \frac{3\alpha}{2}-1)}      E^{-\gamma_0-\alpha}
  \propto  E^{-\gamma_0-\alpha},
\end{equation}
with the injected energy distribution modified due to diffusion. 
In the above expression we assume a spherical symmetry for simplicity with $d = |{\bm r} - {\bm r_s}|$. 
The corresponding energy spectrum of CRs is 
\begin{equation}
   \frac{dN}{dE} =  \iiint d {\bm r} f({\bm r}, E) . 
\end{equation}

\begin{figure*}[ht]
\centering
\subfigure[]{
   \includegraphics[width=8.7cm]{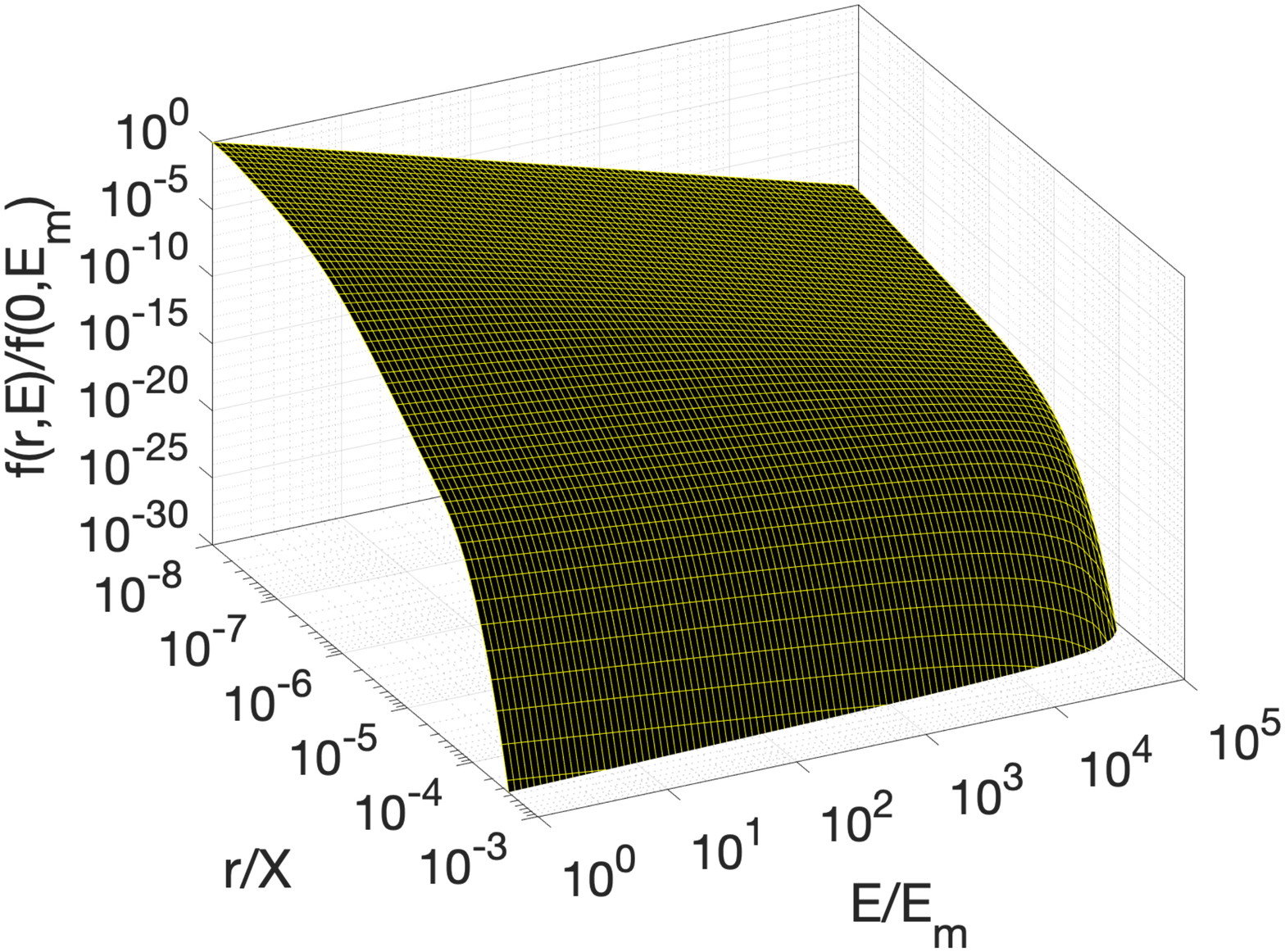}\label{fig: pionmap}}
\subfigure[]{
   \includegraphics[width=8.7cm]{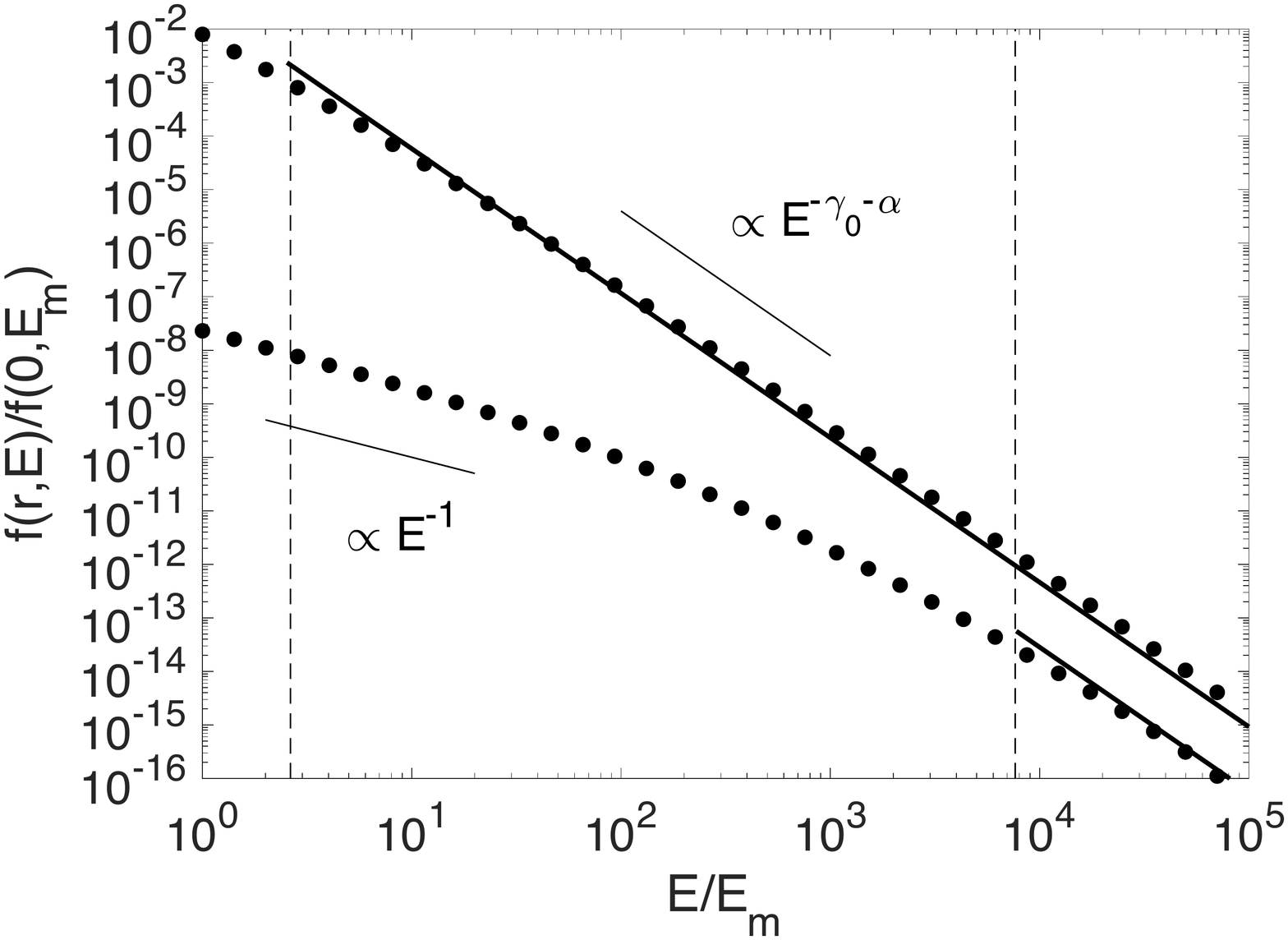}\label{fig: pionspec}}
\caption{(a) The evolution of the normalized distribution function of CRs with the normalized distance from the source and normalized CR energy, 
as a numerical solution to Eq. \eqref{eq: disfun}. 
(b) The normalized energy distribution of CRs at a closer distance with $r/X = 3.4\times10^{-7}$ (upper line) and 
a farther distance with $r/X=5.6\times10^{-6}$ (lower line) from the source. 
The dots and lines represent the numerical solution to Eq. \eqref{eq: disfun} and the analytical approximation at $E>E_d$ (Eq. \eqref{eq: nocool}). 
$E_d$ (Eq. \eqref{eq: enebrk}) is indicated by vertical dashed lines. }
\label{fig: pion}
\end{figure*}

As an illustration, in Fig. \ref{fig: pion} we present the numerical solution to Eq. \eqref{eq: disfun}
by assuming a spherical symmetry. 
We use $f=0$ at $r=X$ as the boundary condition, 
where $r(=d)$ is the radial distance from the source at the center of the sphere, and 
$X$ is the radius of the sphere. 
In fact, $f$ already approaches $0$ before reaching $X$ due to the energy loss (see below). 
Fig. \ref{fig: pionmap} shows the evolution of 
$f(r,E)$ (normalized by its value $f(0,E_m)$ at the source position ($r=0$) and the minimum CR energy $E_m$)
with $r$ (normalized by $X$) and 
E (normalized by $E_m$). 
We note that the parameters adopted here do not correspond to a specific astrophysical object, 
so we only use their normalized values for the illustrative purpose. 
At a closer distance from the source (the upper line in Fig. \ref{fig: pionspec}), 
the energy loss is insignificant for most energies under consideration, and the energy distribution is consistent with 
Eq. \eqref{eq: nocool}.
It is steeper than the injected distribution because of diffusion. 
At a farther distance from the source (the lower line in Fig. \ref{fig: pionspec}), 
as lower-energy CRs diffuse more slowly, they suffer more energy loss 
via hadronic interactions with the surrounding interstellar matter. 
As a result, the injected energy distribution for low-energy CRs cannot be seen. 
The energy scaling is regulated only by energy loss and 
close to Eq. \eqref{eq: lensc} as we analytically expect. 
Higher-energy CRs still follow the energy distribution as in Eq. \eqref{eq: nocool}, with the 
transition energy $E_d$ give by Eq. \eqref{eq: enebrk}. 
Naturally, $E_d$ increases with $d$, 
but decreases with increasing $D_c$. 
With highly suppressed diffusion, 
the energy loss effect can be easily seen within the energy range of interest 
in the vicinity of the CR source.

We find that the energy-dependent diffusion and energy loss together lead to a smoothly broken power law for the CR energy distribution. 
The slope of $dN/dE$ of higher-energy CRs depends on both the injected energy distribution and E dependence of diffusion 
coefficient.

\section{Application to IC 443 and W44}

To exemplify the {mirror} diffusion of CRs in cold interstellar phases with highly {compressible} turbulence, 
we consider the scenario of an SNR interacting with an MC. 
The accelerated CR protons at the SN shock propagate in the 
surrounding MC 
and undergo the {mirror} diffusion when interacting with the {compressible} turbulence.  
To study the gamma-ray emission 
from the dense MC illuminated by the CR protons, 
we adopt the above theoretical model for describing the energy distribution of CRs 
involving both energy-dependent diffusion and energy loss due to pion production. 
We take the parallel diffusion coefficient for {mirror} diffusion of relativistic CRs as 
\begin{equation}\label{eq: appd}
   D
   =10^{23} \Big(\frac{E}{\text{GeV}}\Big)^{10/11} \text{cm}^2\text{s}^{-1}
\end{equation}
based on the calculation in Section \ref{sec: bodiff} and the results in Fig. \ref{fig: diffene}. 
For simplicity, we consider super-Alfv\'{e}nic turbulence
and thus do not distinguish between parallel and perpendicular diffusion in the global frame of reference 
\citep{YL08}. 
We use the above $D$ as the isotropic 
diffusion coefficient, 
but bear in mind this would not apply to sub-Alfv\'{e}nic turbulence. 
For the energy loss due to the inelastic interactions of CR protons with the gas, 
we adopt the model and the value of $\sigma_\text{p,inel}$ in 
\citet{Kel06}.

\begin{figure*}[ht]
\centering
\subfigure[]{
   \includegraphics[width=8.7cm]{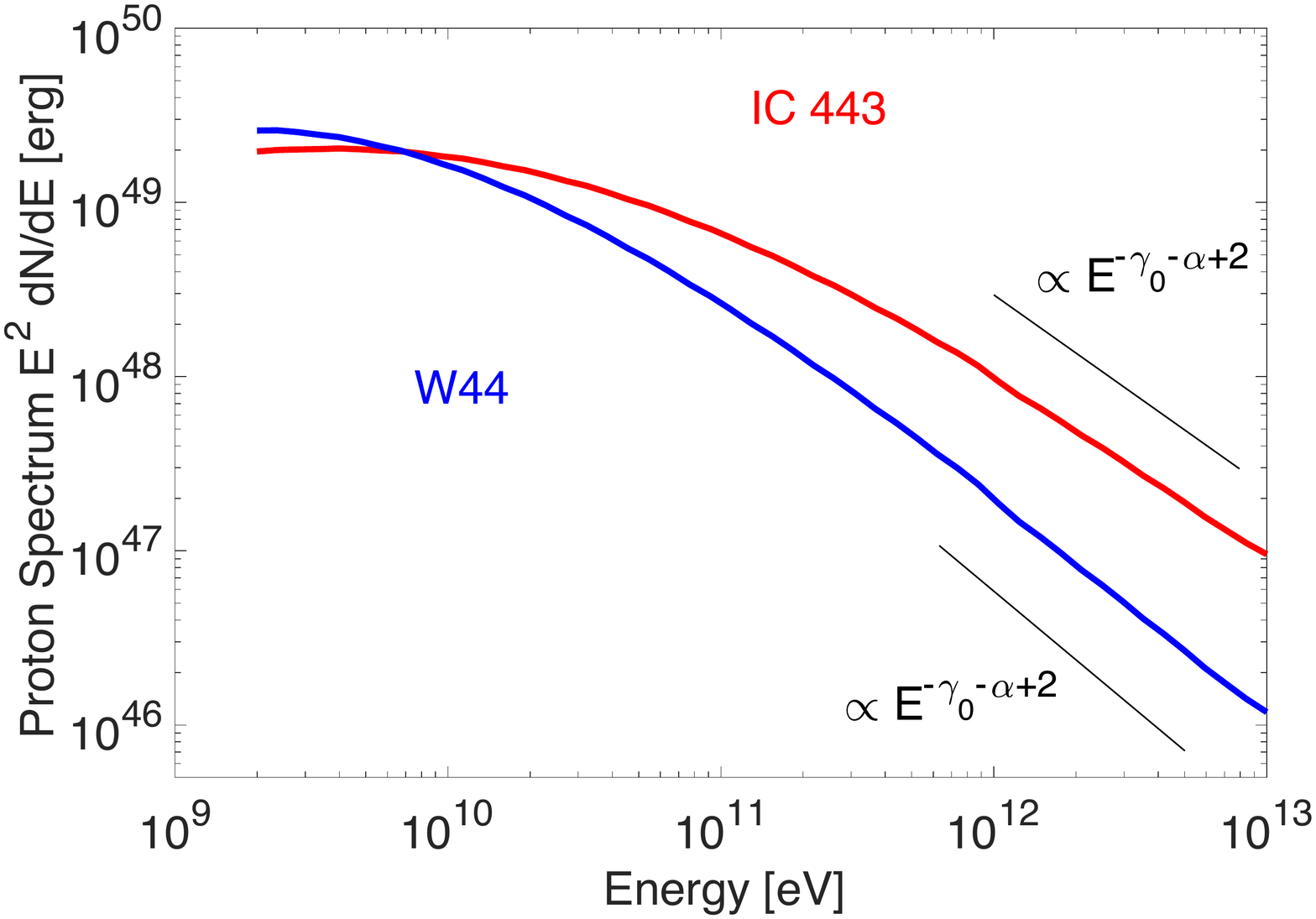}\label{fig: proint}}
\subfigure[]{
   \includegraphics[width=8.7cm]{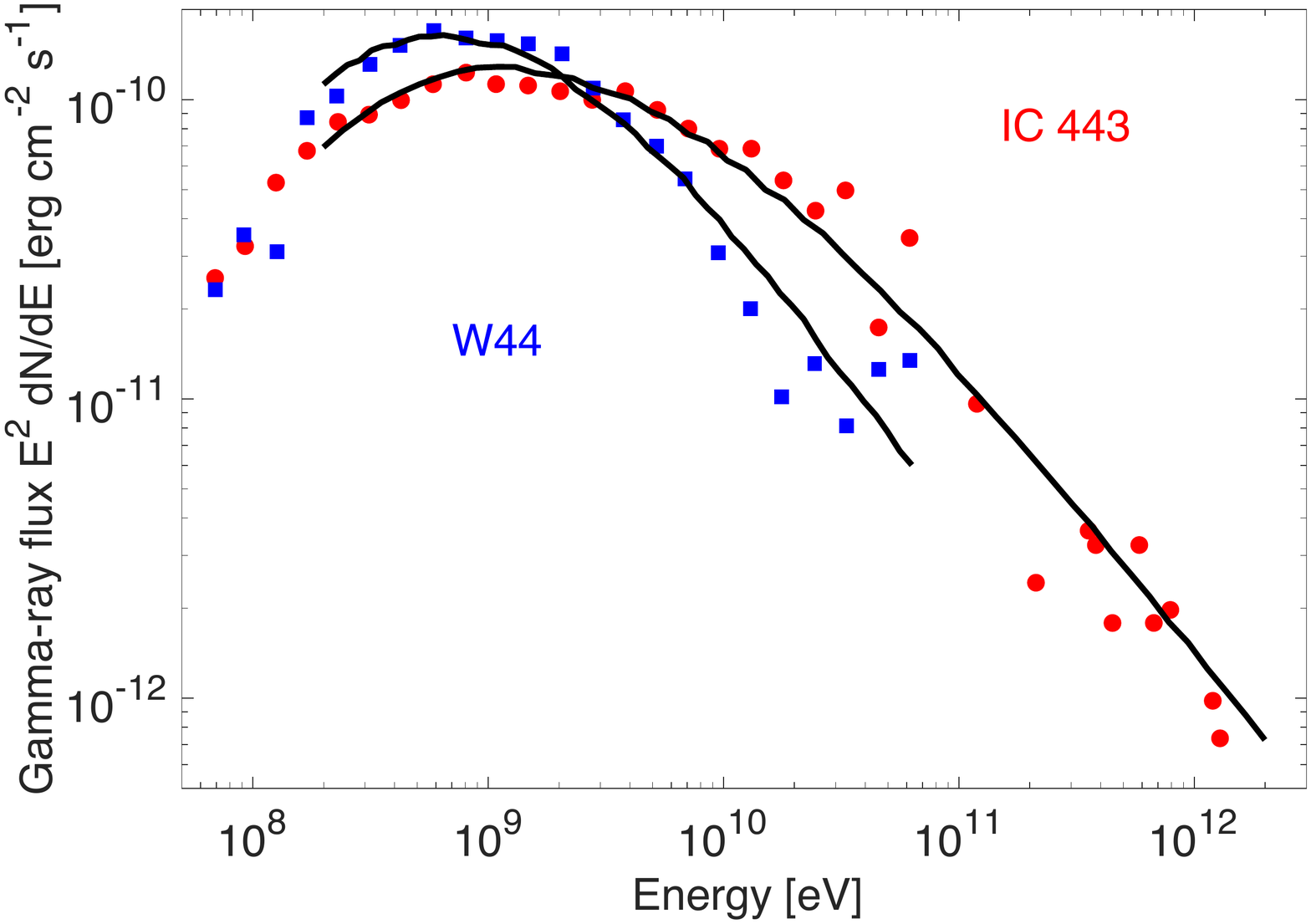}\label{fig: obscom}}
\caption{The modeled CR proton spectra in (a) and gamma-ray spectra in (b) for IC 443 and W44. 
The data points in (b) are taken from 
\citet{Ack13}.
}
\label{fig: modobs}
\end{figure*}

We consider IC 443 and W44, for which the hadronic process mainly contributes to the gamma-ray emission, 
as confirmed based on the characteristic spectral feature expected from neutral pion decay
\citep{Giu11,Ack13}. 
The parameters of these two SNR/MC systems are listed in Table \ref{tab:app}. 
The distance, $B_0$, and $n_H$ are taken from 
\citet{Giu11,Ack13}. 
$B_0$ of tens of $\mu$G and $L$ of tens of pc are typical values for MCs 
(e.g., \citealt{Crut10,Qia18,Ha21}). 
$\gamma_0$ should be in the range $\approx 2.1-2.4$ for shock acceleration 
\citep{St07,Cas11}.
We adopt a larger $\gamma_0$ for W44 due to its particularly steep high-energy gamma-ray spectrum
compared with other middle-aged SNRs
\citep{Card14}. 
Using a spherical symmetry, the CR proton spectrum is 
\begin{equation}
   \frac{dN}{dE} =   
   \int 4\pi r^2 f(r,E) dr, 
\end{equation}
with the range of $r$ for integration provided in Table \ref{tab:app}. 
A broad range of $r$ for IC 443, with a broad range of transition energies corresponding to different 
distances from the source (see Section \ref{sec: discr}), gives rise to an extended bump of the energy spectrum toward low energies. 
For W44 with a higher energy loss rate due to the higher gas density and a narrower spectral bump, 
the range of $r$ is accordingly smaller. 
The resulting CR proton spectra are presented in Fig. \ref{fig: proint}. 
Their corresponding gamma-ray spectra in Fig. \ref{fig: obscom} are calculated 
using the model by 
\citet{Kel06}.
As a comparison, the observational data taken from 
\citet{Ack13}
are also displayed in Fig. \ref{fig: obscom}. 
The value of $Q_c$ is determined to mach the measured gamma-ray flux. 
We note that our approximate expression of $D$ (Eq. \eqref{eq: appd}) for relativistic CRs 
do not apply to CRs with $E \lesssim1$ GeV.

We see that with a suppressed $D$ of {mirror} diffusion in the highly {compressible} turbulence in an MC, 
the energy loss effect is significant and thus modifies the low-energy gamma-ray spectrum. 
Moreover, the steep energy scaling of $D$ of {mirror} diffusion 
explains the steep high-energy gamma-ray spectrum.


\begin{table*}[!htbp]
\renewcommand\arraystretch{1.5}
\centering
\begin{threeparttable}
\caption{Parameters used for modeling the gamma-ray emission from IC 443 and W44.}\label{tab:app} 
  \begin{tabular}{c|c|c|c|c|c|c|c|c}
     \toprule
                &    Distance [kpc] & $B_0$ [$\mu$ G] & $L$ [pc] & $n_H$ [cm$^{-3}$] & $\gamma_0$ & $\alpha$  & Range of $r$ [pc]  & $Q_c$ [erg$^{-1+\gamma_0}$ s$^{-1}$]\\
                \hline
    IC 443 &       $1.5$            & $70$    & $30$ & $20$  & $2.2$ & $10/11$   & $2$-$10$ &  $2.5\times10^{35}$ \\
    W44    &       $2.9$            & $70$    & $30$ & $100$ & $2.4$   &  $10/11$ & $1.6$-$2$   &  $3.0\times10^{36}$ \\
     \bottomrule
    \end{tabular}
 \end{threeparttable}
\end{table*}

\section{Discussion}

The diffusion of CRs in MHD turbulence has the diffusion coefficient $D$
dependent on the properties of MHD turbulence, 
which include, but not limited to the slope of magnetic power spectrum. 
Other important properties, e.g., the scale-dependent anisotropy of Alfv\'{e}n and slow modes
\citep{GS95,LV99,Chan00,YL02,XL20}, 
nonlinear decorrelation of turbulent magnetic fields 
\citep{Ly12,XLb18,Dem19},
can significantly affect the diffusion associated with the pitch angle scattering of CRs.
The {mirror} diffusion of CRs arises from both the mirroring effect of {compressible} modes and 
superdiffusion of magnetic fields induced by Alfv\'{e}n modes, 
and it has dependence on the energy fractions of different modes
(LX21). 
Given the specific turbulence properties in the local environment, 
the corresponding $D$ and its $E$ dependence apply.

As theoretically expected and observationally confirmed, 
the interstellar turbulence has varying properties in different gas phases and regions 
(e.g., \citealt{Armstrong95,Pad99,ElmegreenScalo,CheL10,Chep10,Xup20,Huc21}).
By combining the measurements of $M_s$ and $M_A$  
\citep{Gae11,Toff11,Laz18,Xuy21,Hux21}
that characterize the basic properties of turbulence
and the theoretical model on the turbulence-dependent diffusion of CRs, 
one can obtain a more realistic description of CR diffusion in the multi-phase ISM
and a better interpretation on CR-related observations, e.g., 
gamma-ray spectra.

Here we only consider the case with the {mirror} diffusion dominated by fast modes
by assuming their energy fraction is not small. 
When the energy fraction of fast modes is so small that slow modes dominate the {mirror} diffusion, 
a different D with a different E dependence is expected 
(LX21). 
As the energy fractions of slow and fast modes depend on the local $M_s$ and $M_A$ values
\citep{CL02_PRL,Hu21},
their measurements 
would be necessary for a self-consistent study on CR diffusion 
in SNR/MC systems 
and around other CR sources. 
This would be also helpful for distinguishing between the models using diffusion and 
acceleration of CRs 
(e.g., \citealt{Fang13})
to explain the gamma-ray spectra.


In partially ionized MCs, the ion-neutral collisional damping of fast modes can affect the 
scattering of CRs 
\citep{XLY14,Xuc16,XLb18}. 
The damping effect on {mirror} diffusion of CRs will be addressed in our future studies. 
In addition, if the measured $M_A$ is small,
the anisotropic diffusion in sub-Alfv\'{e}nic turbulence 
\citep{XY13}
should be taken into account when modeling the gamma-ray emission.

The effect of energy loss on modifying the CR energy spectrum in dense MCs 
was earlier discussed in, e.g., 
\citet{Gab07}. 
The main difference is that we use a physically motivated $D$ and specify the shape of CR energy spectrum 
in both energy loss- and diffusion-dominated energy ranges. 
This is also the main difference in comparison with other studies on CR diffusion in SNR/MC systems, 
e.g., \citet{Ohi11,Li12}.


\section{Summary}

The turbulence-dependent diffusion of CRs is spatially inhomogeneous in the multi-phase ISM. 
In the MCs with highly {compressible} turbulence in the vicinity of SNRs,
the {mirror} diffusion of CRs resulting from the mirroring effect of {compressible} magnetic fluctuations 
{and superdiffusion of turbulent magnetic fields} 
is expected. 
As the {mirror} diffusion is more inefficient than other diffusion processes associated with the 
pitch angle scattering of CRs, 
it is likely to be the dominant diffusion process near CR sources, 
accounting for the 
gamma-ray emission from the MCs surrounding SNRs.

Based on the analysis on the distribution function of CRs that undergo both {mirror} diffusion and 
energy loss through hadronic interactions with the ambient material, 
we found a significant energy loss effect on low-energy CRs because of their highly suppressed diffusion in the MCs. 
As a result, the CR energy spectrum has a smoothly broken power law shape, 
with the low-energy part flattened due to the energy loss 
and the high-energy part steepened due to the diffusion. 
This explains the shape of the gamma-ray spectra of IC 443 and W44. 
The diffusion coefficient $D$ of {mirror} diffusion
has a distinctive steep energy scaling
in comparison with other diffusion mechanisms, 
which naturally explains the steep high-energy gamma-ray spectra of SNR/MC systems.

\acknowledgments
S.X. acknowledges the support for 
this work provided by NASA through the NASA Hubble Fellowship grant \# HST-HF2-51473.001-A awarded by the Space Telescope Science Institute, which is operated by the Association of Universities for Research in Astronomy, Incorporated, under NASA contract NAS5-26555. 
{S.X. thanks James Stone for comments that improved the paper.}
\software{MATLAB \citep{MATLAB:2018}}

\appendix 

We provide here the equations used for the numerical calculation of $D_\|$ in Section \ref{sec: bodiff}. 
They are all available in LX21, but still presented here for completeness. 
The {mirroring} rate of fast modes is 
\citep{CesK73,XL20}
\begin{equation}\label{eq: gtraft}
   \Gamma_{b,f}  =   \Big |\frac{1}{\mu} \frac{d\mu}{dt} \Big| =  \frac{v}{2B_0} \frac{1-\mu^2}{\mu}  b_{fk} k  
  = \frac{v}{2L} \frac{\delta B_f^4}{B_0^{4} } \frac{1-\mu^2}{\mu^7}   
\end{equation}
at $\mu>\mu_\text{min,f}$, and 
\begin{equation}\label{eq: trfasmu}
     \Gamma_{b,f}   =   \frac{v}{2B_0} \frac{1-\mu^2}{\mu}  b_{fk}(r_g) r_g^{-1} 
                                                                   =  \frac{v}{2 r_g } \frac{ \delta B_f}{B_0} \Big( \frac{r_g}{L} \Big)^{\frac{1}{4}}  \frac{1-\mu^2}{\mu} 
\end{equation}
at $\mu < \mu_\text{min,f}$.

Slow modes are passively mixed by Alfv\'{e}n modes
\citep{LG01}.
By using the same anisotropic scaling of Alfv\'{e}n modes
\citep{CLV_incomp}, 
\begin{equation}\label{eq: anislber}
    b_{sk} = \delta B_s (k_\perp L)^{-\frac{1}{3}}
               = \delta B_s (k_\| L)^{-\frac{1}{2}},
\end{equation}
where $b_{sk}$ is the magnetic fluctuation of slow modes at $k$, 
and $k_\perp$ and $k_\|$ are the perpendicular and parallel components of $k$
with respect to the local magnetic field 
\citep{LV99},
the {mirroring} rate of slow modes is 
\citep{XL20}
\begin{equation}\label{eq: slbcrlm}
    \Gamma_{b,s}  =  \frac{v}{2B_0} \frac{1-\mu^2}{\mu}  b_{sk} k_\| 
    =  \frac{v}{2 L } \aleph^2_s \frac{1-\mu^2}{\mu^3}     
\end{equation}
at $\mu > \mu_\text{min,s}$, where
\begin{equation}
  \aleph_s  =  \frac{ \delta B_s}{B_0},
\end{equation}
\begin{equation}
   \mu_\text{min,s} = \sqrt{ \frac{b_{sk} (r_g)}{B_0}} = \sqrt{\aleph_s} \Big(\frac{r_g}{ L} \Big)^{\frac{1}{4}},
\end{equation}
and $b_{sk} (r_g)$ is the magnetic fluctuation at $k_\| = 1/ r_g$. 
The {mirroring} rate at $\mu < \mu_\text{min,s}$ is
\begin{equation}\label{eq: trfasmusl}
    \Gamma_{b,s}   = \frac{v}{2B_0} \frac{1-\mu^2}{\mu}  b_{sk} (r_g) r_g^{-1} 
                                                                   = \frac{v}{2 r_g }  \aleph_s \Big(\frac{r_g}{ L} \Big)^{\frac{1}{2}}  \frac{1-\mu^2}{\mu}.    
\end{equation}
For the {mirror} diffusion induced by slow modes, the $\mu$-dependent parallel diffusion coefficient is 
\begin{subnumcases}
     { D_{\|,s} (\mu)=\label{eq: duincst}}
         v\mu k_\|^{-1} =  v L \aleph^{-2}_s \mu^5,  
        ~~~\mu_\text{min,s} <\mu <\mu_c ,\\
         v \mu r_g, ~~~~~~~~~~~~~~~~~~~~~~~~~~~ \mu < \mu_\text{min,s}. \label{eq:duincstsmu}
\end{subnumcases}
Under the assumption of an isotropic pitch angle distribution, 
the corresponding pitch-angle integrated diffusion coefficient is  
\begin{equation}\label{eq: anainc}
   D_{\|,s} \approx \int_0^{\mu_c} D_{\|,s}(\mu) d\mu =\frac{1}{6} vL  \aleph_s^{-2} \mu_c^6.
\end{equation}

For the pitch-angle scattering of CRs, we consider the gyroresonant scattering with the resonance function
in the quasilinear approximation,
\begin{equation}\label{eq: qltreg}
    R = \pi \delta(\omega_k - v_\| k_\| + \Omega )  ,
\end{equation}
where $v_\|$ is the CR speed parallel to the magnetic field, and 
$\omega_k$ is the wave frequency.
The pitch-angle diffusion coefficient for gyroresonant scattering by fast modes is 
\citep{Volk:1975}
\begin{equation}\label{eq: comgyd}
      D_{\mu\mu,f} = C_\mu \int d^3k \frac{k_\|^2}{k^2} [J_1^\prime (x)]^2 I_f(k) R(k)  ,
\end{equation}
with
\begin{equation}\label{eq: cmu}
    C_\mu  = (1 - \mu^2) \frac{\Omega^2}{B_0^2}, ~~ x = \frac{k_\perp v_\perp}{ \Omega}  = \frac{k_\perp }{r_g^{-1} }.
\end{equation}
The energy spectrum is  
\citep{CL02_PRL}
\begin{equation}\label{eq: fsep}
     I_f(k) = C_f k^{-\frac{7}{2}},~~
    C_f = \frac{1}{16 \pi} \delta B_f^2 L^{-\frac{1}{2}}.
\end{equation}

The pitch-angle diffusion coefficients for gyroresonant scattering by 
Alfv\'{e}n and slow modes are 
\citep{Volk:1975}
\begin{equation}\label{eq: oriduav}
     D_{\mu\mu,A} = C_\mu \int d^3k x^{-2} [J_1(x)]^2 I_A(k) R(k),
\end{equation}
and 
\begin{equation}\label{eq: comgyds}
      D_{\mu\mu,s} = C_\mu \int d^3k \frac{k_\|^2}{k^2} [J_1^\prime (x)]^2 I_s(k) R(k)  ,
\end{equation}
where the magnetic energy spectra are  
\citep{CLV_incomp}
\begin{equation}\label{eq: enespa}
     I_A(k) = C_A  k_\perp^{-\frac{10}{3}} \exp{\Bigg(-L^\frac{1}{3}\frac{k_\|}{k_\perp^\frac{2}{3}}\Bigg)},~ 
    C_A = \frac{1}{6 \pi} \delta B_A^2 L^{-\frac{1}{3}}
\end{equation}
for Alfv\'{e}n modes, and 
\begin{equation}\label{eq: slspe}
     I_s(k) = C_s  k_\perp^{-\frac{10}{3}} \exp{\Bigg(-L^\frac{1}{3}\frac{k_\|}{k_\perp^\frac{2}{3}}\Bigg)},~
    C_s = \frac{1}{6 \pi} \delta B_s^2 L^{-\frac{1}{3}}
\end{equation}
for slow modes.

In compressible MHD turbulence, 
we consider the scattering rate with contributions from all three modes, 
\begin{equation}
    \Gamma_{s,\text{tot}} = \frac{2(D_{\mu\mu,A} + D_{\mu\mu,s} + D_{\mu\mu,f})}{\mu^2}.
\end{equation}
$\mu_c$ at the balance between {mirroring} and scattering is determined by equalizing $\Gamma_{s,\text{tot}}$
with $\max[\Gamma_{b,f}, \Gamma_{b,s}]$. 
Then under the assumption of an isotropic pitch angle distribution, 
$D_\|$ for {mirror} diffusion is calculated as 
\begin{equation}
    D_\| = \int_0^{\mu_c} D_{\|}(\mu) d\mu,
\end{equation}
where $D_\|(\mu)$ is the $\mu$-dependent parallel diffusion coefficient for {mirroring} by the 
{compressible} modes with a larger {mirroring} rate.

\bibliographystyle{aasjournal}
\bibliography{xu}

\begin{thebibliography}{}
\expandafter\ifx\csname natexlab\endcsname\relax\def\natexlab#1{#1}\fi
\providecommand{\url}[1]{\href{#1}{#1}}
\providecommand{\dodoi}[1]{doi:~\href{http://doi.org/#1}{\nolinkurl{#1}}}
\providecommand{\doeprint}[1]{\href{http://ascl.net/#1}{\nolinkurl{http://ascl.net/#1}}}
\providecommand{\doarXiv}[1]{\href{https://arxiv.org/abs/#1}{\nolinkurl{https://arxiv.org/abs/#1}}}

\bibitem[{{Abeysekara} {et~al.}(2017){Abeysekara}, {Albert}, {Alfaro},
  {Alvarez}, {{\'A}lvarez}, {Arceo}, {Arteaga-Vel{\'a}zquez}, {Avila Rojas},
  {Ayala Solares}, {Barber}, {Bautista-Elivar}, {Becerril}, {Belmont-Moreno},
  {BenZvi}, {Berley}, {Bernal}, {Braun}, {Brisbois}, {Caballero-Mora},
  {Capistr{\'a}n}, {Carrami{\~n}ana}, {Casanova}, {Castillo}, {Cotti},
  {Cotzomi}, {Couti{\~n}o de Le{\'o}n}, {De Le{\'o}n}, {De la Fuente},
  {Dingus}, {DuVernois}, {D{\'\i}az-V{\'e}lez}, {Ellsworth}, {Engel},
  {Enr{\'\i}quez-Rivera}, {Fiorino}, {Fraija}, {Garc{\'\i}a-Gonz{\'a}lez},
  {Garfias}, {Gerhardt}, {Gonz{\'a}lez Mu{\~n}oz}, {Gonz{\'a}lez}, {Goodman},
  {Hampel-Arias}, {Harding}, {Hern{\'a}ndez}, {Hern{\'a}ndez-Almada}, {Hinton},
  {Hona}, {Hui}, {H{\"u}ntemeyer}, {Iriarte}, {Jardin-Blicq}, {Joshi},
  {Kaufmann}, {Kieda}, {Lara}, {Lauer}, {Lee}, {Lennarz}, {Vargas},
  {Linnemann}, {Longinotti}, {Luis Raya}, {Luna-Garc{\'\i}a}, {L{\'o}pez-Coto},
  {Malone}, {Marinelli}, {Martinez}, {Martinez-Castellanos},
  {Mart{\'\i}nez-Castro}, {Mart{\'\i}nez-Huerta}, {Matthews},
  {Miranda-Romagnoli}, {Moreno}, {Mostaf{\'a}}, {Nellen}, {Newbold}, {Nisa},
  {Noriega-Papaqui}, {Pelayo}, {Pretz}, {P{\'e}rez-P{\'e}rez}, {Ren}, {Rho},
  {Rivi{\`e}re}, {Rosa-Gonz{\'a}lez}, {Rosenberg}, {Ruiz-Velasco}, {Salazar},
  {Salesa Greus}, {Sandoval}, {Schneider}, {Schoorlemmer}, {Sinnis}, {Smith},
  {Springer}, {Surajbali}, {Taboada}, {Tibolla}, {Tollefson}, {Torres},
  {Ukwatta}, {Vianello}, {Weisgarber}, {Westerhoff}, {Wisher}, {Wood},
  {Yapici}, {Yodh}, {Younk}, {Zepeda}, {Zhou}, {Guo}, {Hahn}, {Li}, \&
  {Zhang}}]{Abey17}
{Abeysekara}, A.~U., {Albert}, A., {Alfaro}, R., {et~al.} 2017, Science, 358,
  911, \dodoi{10.1126/science.aan4880}

\bibitem[{{Ackermann} {et~al.}(2013){Ackermann}, {Ajello}, {Allafort},
  {Baldini}, {Ballet}, {Barbiellini}, {Baring}, {Bastieri}, {Bechtol},
  {Bellazzini}, {Blandford}, {Bloom}, {Bonamente}, {Borgland}, {Bottacini},
  {Brandt}, {Bregeon}, {Brigida}, {Bruel}, {Buehler}, {Busetto}, {Buson},
  {Caliandro}, {Cameron}, {Caraveo}, {Casandjian}, {Cecchi}, {{\c{C}}elik},
  {Charles}, {Chaty}, {Chaves}, {Chekhtman}, {Cheung}, {Chiang}, {Chiaro},
  {Cillis}, {Ciprini}, {Claus}, {Cohen-Tanugi}, {Cominsky}, {Conrad}, {Corbel},
  {Cutini}, {D'Ammando}, {de Angelis}, {de Palma}, {Dermer}, {do Couto e
  Silva}, {Drell}, {Drlica-Wagner}, {Falletti}, {Favuzzi}, {Ferrara},
  {Franckowiak}, {Fukazawa}, {Funk}, {Fusco}, {Gargano}, {Germani},
  {Giglietto}, {Giommi}, {Giordano}, {Giroletti}, {Glanzman}, {Godfrey},
  {Grenier}, {Grondin}, {Grove}, {Guiriec}, {Hadasch}, {Hanabata}, {Harding},
  {Hayashida}, {Hayashi}, {Hays}, {Hewitt}, {Hill}, {Hughes}, {Jackson},
  {Jogler}, {J{\'o}hannesson}, {Johnson}, {Kamae}, {Kataoka}, {Katsuta},
  {Kn{\"o}dlseder}, {Kuss}, {Lande}, {Larsson}, {Latronico}, {Lemoine-Goumard},
  {Longo}, {Loparco}, {Lovellette}, {Lubrano}, {Madejski}, {Massaro}, {Mayer},
  {Mazziotta}, {McEnery}, {Mehault}, {Michelson}, {Mignani}, {Mitthumsiri},
  {Mizuno}, {Moiseev}, {Monzani}, {Morselli}, {Moskalenko}, {Murgia},
  {Nakamori}, {Nemmen}, {Nuss}, {Ohno}, {Ohsugi}, {Omodei}, {Orienti},
  {Orlando}, {Ormes}, {Paneque}, {Perkins}, {Pesce-Rollins}, {Piron}, {Pivato},
  {Rain{\`o}}, {Rando}, {Razzano}, {Razzaque}, {Reimer}, {Reimer}, {Ritz},
  {Romoli}, {S{\'a}nchez-Conde}, {Schulz}, {Sgr{\`o}}, {Simeon}, {Siskind},
  {Smith}, {Spandre}, {Spinelli}, {Stecker}, {Strong}, {Suson}, {Tajima},
  {Takahashi}, {Takahashi}, {Tanaka}, {Thayer}, {Thayer}, {Thompson},
  {Thorsett}, {Tibaldo}, {Tibolla}, {Tinivella}, {Troja}, {Uchiyama}, {Usher},
  {Vandenbroucke}, {Vasileiou}, {Vianello}, {Vitale}, {Waite}, {Werner},
  {Winer}, {Wood}, {Wood}, {Yamazaki}, {Yang}, \& {Zimmer}}]{Ack13}
{Ackermann}, M., {Ajello}, M., {Allafort}, A., {et~al.} 2013, Science, 339,
  807, \dodoi{10.1126/science.1231160}

\bibitem[{{Armstrong} {et~al.}(1995){Armstrong}, {Rickett}, \&
  {Spangler}}]{Armstrong95}
{Armstrong}, J.~W., {Rickett}, B.~J., \& {Spangler}, S.~R. 1995, \apj, 443,
  209, \dodoi{10.1086/175515}

\bibitem[{{Axford} {et~al.}(1977){Axford}, {Leer}, \& {Skadron}}]{Axf77}
{Axford}, W.~I., {Leer}, E., \& {Skadron}, G. 1977, in International Cosmic Ray
  Conference, Vol.~11, International Cosmic Ray Conference, 132

\bibitem[{{Bally} {et~al.}(1987){Bally}, {Langer}, {Stark}, \&
  {Wilson}}]{Bal87}
{Bally}, J., {Langer}, W.~D., {Stark}, A.~A., \& {Wilson}, R.~W. 1987, \apjl,
  312, L45, \dodoi{10.1086/184817}

\bibitem[{{Beresnyak} {et~al.}(2011){Beresnyak}, {Yan}, \&
  {Lazarian}}]{BYL2011}
{Beresnyak}, A., {Yan}, H., \& {Lazarian}, A. 2011, \apj, 728, 60,
  \dodoi{10.1088/0004-637X/728/1/60}

\bibitem[{{Blasi} \& {Amato}(2012)}]{Blaa12}
{Blasi}, P., \& {Amato}, E. 2012, Journal of Cosmology and Astroparticle
  Physics, 2012, 010, \dodoi{10.1088/1475-7516/2012/01/010}

\bibitem[{{Blasi} {et~al.}(2012){Blasi}, {Amato}, \& {Serpico}}]{Bla12}
{Blasi}, P., {Amato}, E., \& {Serpico}, P.~D. 2012, \prl, 109, 061101,
  \dodoi{10.1103/PhysRevLett.109.061101}

\bibitem[{{Brunetti} \& {Lazarian}(2007)}]{Brunetti_Laz}
{Brunetti}, G., \& {Lazarian}, A. 2007, \mnras, 378, 245,
  \dodoi{10.1111/j.1365-2966.2007.11771.x}

\bibitem[{{Brunetti} \& {Lazarian}(2011)}]{BruLaz11}
---. 2011, \mnras, 412, 817, \dodoi{10.1111/j.1365-2966.2010.17937.x}

\bibitem[{{Cardillo} {et~al.}(2014){Cardillo}, {Tavani}, {Giuliani},
  {Yoshiike}, {Sano}, {Fukuda}, {Fukui}, {Castelletti}, \& {Dubner}}]{Card14}
{Cardillo}, M., {Tavani}, M., {Giuliani}, A., {et~al.} 2014, \aap, 565, A74,
  \dodoi{10.1051/0004-6361/201322685}

\bibitem[{{Castellina} \& {Donato}(2011)}]{Cas11}
{Castellina}, A., \& {Donato}, F. 2011, arXiv:1110.2981, arXiv:1110.2981.
\newblock \doarXiv{1110.2981}

\bibitem[{{Ceccarelli} {et~al.}(2011){Ceccarelli}, {Hily-Blant}, {Montmerle},
  {Dubus}, {Gallant}, \& {Fiasson}}]{Cec11}
{Ceccarelli}, C., {Hily-Blant}, P., {Montmerle}, T., {et~al.} 2011, \apjl, 740,
  L4, \dodoi{10.1088/2041-8205/740/1/L4}

\bibitem[{{Cesarsky} \& {Kulsrud}(1973)}]{CesK73}
{Cesarsky}, C.~J., \& {Kulsrud}, R.~M. 1973, ApJ, 185, 153

\bibitem[{{Chandran}(2000)}]{Chan00}
{Chandran}, B.~D.~G. 2000, Physical Review Letters, 85, 4656,
  \dodoi{10.1103/PhysRevLett.85.4656}

\bibitem[{{Chepurnov} \& {Lazarian}(2010)}]{CheL10}
{Chepurnov}, A., \& {Lazarian}, A. 2010, \apj, 710, 853,
  \dodoi{10.1088/0004-637X/710/1/853}

\bibitem[{{Chepurnov} {et~al.}(2010){Chepurnov}, {Lazarian},
  {Stanimirovi{\'c}}, {Heiles}, \& {Peek}}]{Chep10}
{Chepurnov}, A., {Lazarian}, A., {Stanimirovi{\'c}}, S., {Heiles}, C., \&
  {Peek}, J.~E.~G. 2010, \apj, 714, 1398, \dodoi{10.1088/0004-637X/714/2/1398}

\bibitem[{{Cho} \& {Lazarian}(2002)}]{CL02_PRL}
{Cho}, J., \& {Lazarian}, A. 2002, Physical Review Letters, 88, 245001,
  \dodoi{10.1103/PhysRevLett.88.245001}

\bibitem[{{Cho} \& {Lazarian}(2003)}]{CL03}
---. 2003, \mnras, 345, 325, \dodoi{10.1046/j.1365-8711.2003.06941.x}

\bibitem[{{Cho} {et~al.}(2002){Cho}, {Lazarian}, \& {Vishniac}}]{CLV_incomp}
{Cho}, J., {Lazarian}, A., \& {Vishniac}, E.~T. 2002, \apj, 564, 291,
  \dodoi{10.1086/324186}

\bibitem[{{Cho} \& {Vishniac}(2000)}]{CV00}
{Cho}, J., \& {Vishniac}, E.~T. 2000, \apj, 539, 273, \dodoi{10.1086/309213}

\bibitem[{{Crutcher} {et~al.}(2010){Crutcher}, {Wandelt}, {Heiles},
  {Falgarone}, \& {Troland}}]{Crut10}
{Crutcher}, R.~M., {Wandelt}, B., {Heiles}, C., {Falgarone}, E., \& {Troland},
  T.~H. 2010, \apj, 725, 466, \dodoi{10.1088/0004-637X/725/1/466}

\bibitem[{{Demidem} {et~al.}(2019){Demidem}, {Lemoine}, \& {Casse}}]{Dem19}
{Demidem}, C., {Lemoine}, M., \& {Casse}, F. 2019, arXiv:1909.12885,
  arXiv:1909.12885.
\newblock \doarXiv{1909.12885}

\bibitem[{{Di Sciascio}(2019)}]{DiS19}
{Di Sciascio}, G. 2019, in Journal of Physics Conference Series, Vol. 1263,
  Journal of Physics Conference Series, 012003,
  \dodoi{10.1088/1742-6596/1263/1/012003}

\bibitem[{{Elmegreen} \& {Scalo}(2004)}]{ElmegreenScalo}
{Elmegreen}, B.~G., \& {Scalo}, J. 2004, \araa, 42, 211,
  \dodoi{10.1146/annurev.astro.41.011802.094859}

\bibitem[{{Evoli} {et~al.}(2018){Evoli}, {Linden}, \& {Morlino}}]{Evo18}
{Evoli}, C., {Linden}, T., \& {Morlino}, G. 2018, \prd, 98, 063017,
  \dodoi{10.1103/PhysRevD.98.063017}

\bibitem[{{Eyink} {et~al.}(2011){Eyink}, {Lazarian}, \& {Vishniac}}]{Eyink2011}
{Eyink}, G.~L., {Lazarian}, A., \& {Vishniac}, E.~T. 2011, \apj, 743, 51,
  \dodoi{10.1088/0004-637X/743/1/51}

\bibitem[{{Fang} {et~al.}(2013){Fang}, {Yu}, {Zhu}, \& {Zhang}}]{Fang13}
{Fang}, J., {Yu}, H., {Zhu}, B.-T., \& {Zhang}, L. 2013, \mnras, 435, 570,
  \dodoi{10.1093/mnras/stt1318}

\bibitem[{{Federrath} {et~al.}(2011){Federrath}, {Chabrier}, {Schober},
  {Banerjee}, {Klessen}, \& {Schleicher}}]{Fede11}
{Federrath}, C., {Chabrier}, G., {Schober}, J., {et~al.} 2011, Physical Review
  Letters, 107, 114504, \dodoi{10.1103/PhysRevLett.107.114504}

\bibitem[{{Funk}(2017)}]{Fun17}
{Funk}, S. 2017, {High-Energy Gamma Rays from Supernova Remnants}, ed. A.~W.
  {Alsabti} \& P.~{Murdin}, 1737, \dodoi{10.1007/978-3-319-21846-5\_12}

\bibitem[{{Gabici} {et~al.}(2007){Gabici}, {Aharonian}, \& {Blasi}}]{Gab07}
{Gabici}, S., {Aharonian}, F.~A., \& {Blasi}, P. 2007, \apss, 309, 365,
  \dodoi{10.1007/s10509-007-9427-6}

\bibitem[{{Gaensler} {et~al.}(2011){Gaensler}, {Haverkorn}, {Burkhart},
  {Newton-McGee}, {Ekers}, {Lazarian}, {McClure-Griffiths}, {Robishaw},
  {Dickey}, \& {Green}}]{Gae11}
{Gaensler}, B.~M., {Haverkorn}, M., {Burkhart}, B., {et~al.} 2011, \nat, 478,
  214, \dodoi{10.1038/nature10446}

\bibitem[{{Ginzburg} \& {Syrovatskii}(1964)}]{Gin64}
{Ginzburg}, V.~L., \& {Syrovatskii}, S.~I. 1964, {The Origin of Cosmic Rays}
  (Pergamon Press, Oxford)

\bibitem[{{Giuliani} {et~al.}(2011){Giuliani}, {Cardillo}, {Tavani}, {Fukui},
  {Yoshiike}, {Torii}, {Dubner}, {Castelletti}, {Barbiellini}, {Bulgarelli},
  {Caraveo}, {Costa}, {Cattaneo}, {Chen}, {Contessi}, {Del Monte},
  {Donnarumma}, {Evangelista}, {Feroci}, {Gianotti}, {Lazzarotto}, {Lucarelli},
  {Longo}, {Marisaldi}, {Mereghetti}, {Pacciani}, {Pellizzoni}, {Piano},
  {Picozza}, {Pittori}, {Pucella}, {Rapisarda}, {Rappoldi}, {Sabatini},
  {Soffitta}, {Striani}, {Trifoglio}, {Trois}, {Vercellone}, {Verrecchia},
  {Vittorini}, {Colafrancesco}, {Giommi}, \& {Bignami}}]{Giu11}
{Giuliani}, A., {Cardillo}, M., {Tavani}, M., {et~al.} 2011, \apjl, 742, L30,
  \dodoi{10.1088/2041-8205/742/2/L30}

\bibitem[{{Goldreich} \& {Sridhar}(1995)}]{GS95}
{Goldreich}, P., \& {Sridhar}, S. 1995, \apj, 438, 763, \dodoi{10.1086/175121}

\bibitem[{{Gratton}(1972)}]{Gra72}
{Gratton}, L. 1972, \apss, 16, 81, \dodoi{10.1007/BF00643094}

\bibitem[{{Grie{\ss}meier} {et~al.}(2015){Grie{\ss}meier}, {Tabataba-Vakili},
  {Stadelmann}, {Grenfell}, \& {Atri}}]{Grie15}
{Grie{\ss}meier}, J.~M., {Tabataba-Vakili}, F., {Stadelmann}, A., {Grenfell},
  J.~L., \& {Atri}, D. 2015, \aap, 581, A44,
  \dodoi{10.1051/0004-6361/201425451}

\bibitem[{{Ha} {et~al.}(2021){Ha}, {Li}, {Xu}, {Kounkel}, \& {Li}}]{Ha21}
{Ha}, T., {Li}, Y., {Xu}, S., {Kounkel}, M., \& {Li}, H. 2021, \apjl, 907, L40,
  \dodoi{10.3847/2041-8213/abd8c9}

\bibitem[{{Hu} {et~al.}(2021{\natexlab{a}}){Hu}, {Lazarian}, \& {Wang}}]{Huc21}
{Hu}, Y., {Lazarian}, A., \& {Wang}, Q.~D. 2021{\natexlab{a}}, arXiv e-prints,
  arXiv:2105.03605.
\newblock \doarXiv{2105.03605}

\bibitem[{{Hu} {et~al.}(2021{\natexlab{b}}){Hu}, {Xu}, \& {Lazarian}}]{Hux21}
{Hu}, Y., {Xu}, S., \& {Lazarian}, A. 2021{\natexlab{b}}, \apj, 911, 37,
  \dodoi{10.3847/1538-4357/abea18}

\bibitem[{{Hu}(2021)}]{Hu21}
{Hu}, Y. e.~a. 2021, in prep

\bibitem[{{Indriolo} {et~al.}(2010){Indriolo}, {Blake}, {Goto}, {Usuda}, {Oka},
  {Geballe}, {Fields}, \& {McCall}}]{Ind10}
{Indriolo}, N., {Blake}, G.~A., {Goto}, M., {et~al.} 2010, \apj, 724, 1357,
  \dodoi{10.1088/0004-637X/724/2/1357}

\bibitem[{{Ipavich}(1975)}]{Ipa75}
{Ipavich}, F.~M. 1975, \apj, 196, 107, \dodoi{10.1086/153397}

\bibitem[{{Kelner} {et~al.}(2006){Kelner}, {Aharonian}, \& {Bugayov}}]{Kel06}
{Kelner}, S.~R., {Aharonian}, F.~A., \& {Bugayov}, V.~V. 2006, \prd, 74,
  034018, \dodoi{10.1103/PhysRevD.74.034018}

\bibitem[{{Lazarian}(2006)}]{Lazarian06}
{Lazarian}, A. 2006, \apjl, 645, L25, \dodoi{10.1086/505796}

\bibitem[{{Lazarian} \& {Vishniac}(1999)}]{LV99}
{Lazarian}, A., \& {Vishniac}, E.~T. 1999, \apj, 517, 700,
  \dodoi{10.1086/307233}

\bibitem[{{Lazarian} \& {Xu}(2021)}]{LX21}
{Lazarian}, A., \& {Xu}, S. 2021, submitted to ApJ

\bibitem[{{Lazarian} \& {Yan}(2014)}]{LY14}
{Lazarian}, A., \& {Yan}, H. 2014, \apj, 784, 38,
  \dodoi{10.1088/0004-637X/784/1/38}

\bibitem[{{Lazarian} {et~al.}(2018){Lazarian}, {Yuen}, {Ho}, {Chen},
  {Lazarian}, {Lu}, {Yang}, \& {Hu}}]{Laz18}
{Lazarian}, A., {Yuen}, K.~H., {Ho}, K.~W., {et~al.} 2018, \apj, 865, 46,
  \dodoi{10.3847/1538-4357/aad7ff}

\bibitem[{{Li} \& {Chen}(2012)}]{Li12}
{Li}, H., \& {Chen}, Y. 2012, \mnras, 421, 935,
  \dodoi{10.1111/j.1365-2966.2012.20270.x}

\bibitem[{{Lim} {et~al.}(2020){Lim}, {Cho}, \& {Yoon}}]{Lim20}
{Lim}, J., {Cho}, J., \& {Yoon}, H. 2020, \apj, 893, 75,
  \dodoi{10.3847/1538-4357/ab8066}

\bibitem[{{Lithwick} \& {Goldreich}(2001)}]{LG01}
{Lithwick}, Y., \& {Goldreich}, P. 2001, \apj, 562, 279, \dodoi{10.1086/323470}

\bibitem[{{Lynn} {et~al.}(2012){Lynn}, {Parrish}, {Quataert}, \&
  {Chandran}}]{Ly12}
{Lynn}, J.~W., {Parrish}, I.~J., {Quataert}, E., \& {Chandran}, B.~D.~G. 2012,
  \apj, 758, 78, \dodoi{10.1088/0004-637X/758/2/78}

\bibitem[{{Mac Low} \& {Klessen}(2004)}]{MacL04}
{Mac Low}, M.-M., \& {Klessen}, R.~S. 2004, Reviews of Modern Physics, 76, 125,
  \dodoi{10.1103/RevModPhys.76.125}

\bibitem[{{Mannheim} \& {Schlickeiser}(1994)}]{ManS94}
{Mannheim}, K., \& {Schlickeiser}, R. 1994, \aap, 286, 983.
\newblock \doarXiv{astro-ph/9402042}

\bibitem[{MATLAB(2018)}]{MATLAB:2018}
MATLAB. 2018, 9.7.0.1190202 (R2019b) (Natick, Massachusetts: The MathWorks
  Inc.)

\bibitem[{{Neronov} {et~al.}(2012){Neronov}, {Semikoz}, \& {Taylor}}]{Nero12}
{Neronov}, A., {Semikoz}, D.~V., \& {Taylor}, A.~M. 2012, \prl, 108, 051105,
  \dodoi{10.1103/PhysRevLett.108.051105}

\bibitem[{{Ohira} {et~al.}(2011){Ohira}, {Murase}, \& {Yamazaki}}]{Ohi11}
{Ohira}, Y., {Murase}, K., \& {Yamazaki}, R. 2011, \mnras, 410, 1577,
  \dodoi{10.1111/j.1365-2966.2010.17539.x}

\bibitem[{{Padoan} {et~al.}(1999){Padoan}, {Bally}, {Billawala}, {Juvela}, \&
  {Nordlund}}]{Pad99}
{Padoan}, P., {Bally}, J., {Billawala}, Y., {Juvela}, M., \& {Nordlund},
  {\r{A}}. 1999, \apj, 525, 318, \dodoi{10.1086/307864}

\bibitem[{{Padovani} {et~al.}(2018){Padovani}, {Ivlev}, {Galli}, \&
  {Caselli}}]{Padd18}
{Padovani}, M., {Ivlev}, A.~V., {Galli}, D., \& {Caselli}, P. 2018, \aap, 614,
  A111, \dodoi{10.1051/0004-6361/201732202}

\bibitem[{{Potgieter}(2013)}]{Pot13}
{Potgieter}, M.~S. 2013, Living Reviews in Solar Physics, 10, 3,
  \dodoi{10.12942/lrsp-2013-3}

\bibitem[{{Qian} {et~al.}(2018){Qian}, {Li}, {Gao}, {Xu}, \& {Pan}}]{Qia18}
{Qian}, L., {Li}, D., {Gao}, Y., {Xu}, H., \& {Pan}, Z. 2018, \apj, 864, 116,
  \dodoi{10.3847/1538-4357/aad780}

\bibitem[{{Schlickeiser} {et~al.}(2016){Schlickeiser}, {Caglar}, \&
  {Lazarian}}]{Schlk16}
{Schlickeiser}, R., {Caglar}, M., \& {Lazarian}, A. 2016, \apj, 824, 89,
  \dodoi{10.3847/0004-637X/824/2/89}

\bibitem[{{Semenov} {et~al.}(2021){Semenov}, {Kravtsov}, \&
  {Caprioli}}]{Seme21}
{Semenov}, V.~A., {Kravtsov}, A.~V., \& {Caprioli}, D. 2021, \apj, 910, 126,
  \dodoi{10.3847/1538-4357/abe2a6}

\bibitem[{{Strong} {et~al.}(2007){Strong}, {Moskalenko}, \& {Ptuskin}}]{St07}
{Strong}, A.~W., {Moskalenko}, I.~V., \& {Ptuskin}, V.~S. 2007, Annual Review
  of Nuclear and Particle Science, 57, 285,
  \dodoi{10.1146/annurev.nucl.57.090506.123011}

\bibitem[{{Syrovatskii}(1959)}]{Syr59}
{Syrovatskii}, S.~I. 1959, \sovast, 3, 22

\bibitem[{{Tofflemire} {et~al.}(2011){Tofflemire}, {Burkhart}, \&
  {Lazarian}}]{Toff11}
{Tofflemire}, B.~M., {Burkhart}, B., \& {Lazarian}, A. 2011, \apj, 736, 60,
  \dodoi{10.1088/0004-637X/736/1/60}

\bibitem[{{Torres} {et~al.}(2010){Torres}, {Marrero}, \& {de Cea Del
  Pozo}}]{Tor10}
{Torres}, D.~F., {Marrero}, A. Y.~R., \& {de Cea Del Pozo}, E. 2010, \mnras,
  408, 1257, \dodoi{10.1111/j.1365-2966.2010.17205.x}

\bibitem[{{Vazquez-Semadeni} {et~al.}(2000){Vazquez-Semadeni}, {Ostriker},
  {Passot}, {Gammie}, \& {Stone}}]{Vaz00}
{Vazquez-Semadeni}, E., {Ostriker}, E.~C., {Passot}, T., {Gammie}, C.~F., \&
  {Stone}, J.~M. 2000, in Protostars and Planets IV, ed. V.~{Mannings}, A.~P.
  {Boss}, \& S.~S. {Russell}, 3.
\newblock \doarXiv{astro-ph/9903066}

\bibitem[{{Voelk}(1975)}]{Volk:1975}
{Voelk}, H.~J. 1975, Reviews of Geophysics and Space Physics, 13, 547,
  \dodoi{10.1029/RG013i004p00547}

\bibitem[{{Xu} \& {Hu}(2021)}]{Xuy21}
{Xu}, S., \& {Hu}, Y. 2021, \apj, 910, 88, \dodoi{10.3847/1538-4357/abe403}

\bibitem[{{Xu} {et~al.}(2019){Xu}, {Ji}, \& {Lazarian}}]{XJL19}
{Xu}, S., {Ji}, S., \& {Lazarian}, A. 2019, \apj, 878, 157,
  \dodoi{10.3847/1538-4357/ab21be}

\bibitem[{{Xu} \& {Lazarian}(2018)}]{XLb18}
{Xu}, S., \& {Lazarian}, A. 2018, \apj, 868, 36,
  \dodoi{10.3847/1538-4357/aae840}

\bibitem[{{Xu} \& {Lazarian}(2020)}]{XL20}
---. 2020, \apj, 894, 63, \dodoi{10.3847/1538-4357/ab8465}

\bibitem[{{Xu} {et~al.}(2015){Xu}, {Lazarian}, \& {Yan}}]{XLY14}
{Xu}, S., {Lazarian}, A., \& {Yan}, H. 2015, \apj, 810, 44,
  \dodoi{10.1088/0004-637X/810/1/44}

\bibitem[{{Xu} \& {Yan}(2013)}]{XY13}
{Xu}, S., \& {Yan}, H. 2013, \apj, 779, 140,
  \dodoi{10.1088/0004-637X/779/2/140}

\bibitem[{{Xu} {et~al.}(2016){Xu}, {Yan}, \& {Lazarian}}]{Xuc16}
{Xu}, S., {Yan}, H., \& {Lazarian}, A. 2016, \apj, 826, 166,
  \dodoi{10.3847/0004-637X/826/2/166}

\bibitem[{{Xu} \& {Zhang}(2016)}]{XuZ16}
{Xu}, S., \& {Zhang}, B. 2016, \apj, 824, 113,
  \dodoi{10.3847/0004-637X/824/2/113}

\bibitem[{{Xu} \& {Zhang}(2017)}]{XuZ17}
---. 2017, \apj, 835, 2, \dodoi{10.3847/1538-4357/835/1/2}

\bibitem[{{Xu} \& {Zhang}(2020)}]{Xup20}
---. 2020, \apj, 905, 159, \dodoi{10.3847/1538-4357/abc69f}

\bibitem[{{Yan} \& {Lazarian}(2002)}]{YL02}
{Yan}, H., \& {Lazarian}, A. 2002, Physical Review Letters, 89, B1102+,
  \dodoi{10.1103/PhysRevLett.89.281102}

\bibitem[{{Yan} \& {Lazarian}(2004)}]{YL04}
---. 2004, \apj, 614, 757, \dodoi{10.1086/423733}

\bibitem[{{Yan} \& {Lazarian}(2008)}]{YL08}
---. 2008, \apj, 673, 942, \dodoi{10.1086/524771}

\end{thebibliography}

\end{document}